\documentstyle[preprint,aps]{revtex}

\begin{document}

\input{epsf}
\draft
%\preprint{{\bf MITH 98/9}\\ \\ }
\title{ What hides at the center of the Sun ?}

\author{{\bf G. Preparata\footnote{On sabbatical leave at ENEA, 
Centro Ricerche di Frascati, Frascati (Italy)}}}
\address{
Dipartimento di Fisica, Universit\`a di Milano\\ 
INFN, Sezione di Milano, Italy. e-mail: Preparata@mi.infn.it
}
%\date{October 26, 1998}@
\maketitle
\begin{abstract}
The Standard Solar Model (SSM) and its basic assumption of an equation 
of state of perfect gases for the plasmas of its interior is analyzed 
within the new theoretical framework of QED coherent states. We find 
that for $\frac{r}{r_{\odot}}\leq 0.3$ the SSM solution is unstable 
against collapse to a dense, {\it cold} coherent core and as a result 
a large suppression of $B^{8}$-neutrinos is obtained, apparently solving 
the long standing solar neutrino problem.
\end{abstract}

%\pacs{PACS: 13.85 Q, 24.30 C, 25.70 E}

\vfill\eject

The Sun, so crucial to our human destiny, is but an average star, one 
rather anonimous among the $10^{20}$ or so that populate our Universe. 
And, as a typical representative of such huge population, being so 
close to us, it has been the object of extensive studies both 
experimental and theoretical such that most astrophysicists are today 
convinced that indeed we do know what is going on inside it. In this 
way a rather tight model of the Sun's interior has been built - the 
Standard Solar Model (SSM) - that accounts in a consistent fashion 
for the most diverse features of this average star, from its 
luminosity to the pletora of nuclear fusion reactions that are 
responsible for its long safe burning. All good and in order then, 
not quite: in the last thirty years \textit{a small cloud has appeared 
in this serene, luminous sky}: the solar neutrino problem\footnote{For 
a nice, thorough account of this fascinating problem and for an 
extensive bibliography see the book of Ref \cite{Bahcall}}.
But the faith in the SSM has in no way faltered, rather the puzzle is 
now generally believed to be solved by the old Pontecorvo idea of \textit{ 
neutrino mixing}, a subtle phenomenon that would reproduce in the 
lepton sector of the Standard Model of particle interactions the well 
established mixing of the quark sector. However, in spite of the 
remarkable effort that high energy physicists have made in the last 
decades to obtain a consistent mixing scheme, to this day no reliable 
solution is known, and the search continues.\\
In this paper, instead, we shall try and submit to a critical 
evaluation the SSM itself with two main aims: first to assess whether 
the basic assumption of the equation of state of a perfect gas, that 
appears so natural is indeed so, and second whether a new consistent 
description of the Sun's interior exists that, leading to different 
basic parameters (pressure, temperature, density), solves the solar 
neutrino problem.\\
The framework in which such analysis shall be performed is that of  
\textit{ QED Coherence}, expounded at length in a recent book 
\cite{Preparata} that systematically describes a new class of 
solutions of QED dynamical equations in condensed matter, whose 
existence has been confirmed by several Authors, and in particular by 
the well known condensed matter theorist C.P. Enz \cite{Enz}. As this 
new approach to condensed matter physics is not widely known, let us 
first review its main points concerning the plasmas of electrons, 
protons and several other nuclei that are expected to crowd the Sun's 
center. The general theory of coherent plasmas, described in Chap. 5 
of Ref\cite{Enz}, leads to the following conclusions:

\begin{description}
\item 
(i) \ at zero temperature the gaseous state of each plasma, 
characterized by its plasma
frequency \footnote{Throughout this paper we shall use the 
\textit{natural units system} in which $\hbar = c = k = 1$} 

\begin{equation}
\omega_{p} = 
\frac{e}{m^{\frac{1}{2}}}\left(\frac{N}{V}\right)^{\frac{1}{2}};
\label{eq:1} 
\end{equation}

is unstable against a transition to a coherent laser-like state, 
where all particles of the plasma oscillate in phase with a coherent 
electromagnetic field, which thus realizes an energy gain; 

	\item  е
(ii)\ the energy gain per particle, the gap $\Delta$, is given by 
($\rho = \left(\frac{N}{V}\right)$ is the number density)

\begin{equation}
\Delta = 2.95e^{2}\rho^{\frac{1}{3}} ;е
\label{eq:2} 
\end{equation}
 
	\item  е
(iii)\ for $T\neq 0$, due to thermal fluctuations an incoherent plasma 
develops, which coexists with the coherent one, whose fraction $F_{i}$  
for a spin$-\frac{1}{2}$ fermion is given by

\begin{equation}
F_{i} = 
\frac{2}{\rho}\int_{p_{F}}\frac{d^{3}p}{(2\pi)^{3}}\frac{1}{exp(\frac{E-E_{F}+\Delta}{T})+1}
\simeq \frac{1}{\pi}\frac{mT}{\rho^{\frac{2}{3}}}exp(-\frac{\Delta}{T})  ;е
\label{eq:3} 
\end{equation}

where $p_{F}е$ is the Fermi momentum and $E_{F}=\frac{p^{2}_{F}ее}{2m}$ 
is the Fermi energy;

      \item
 (iv)\ in order to determine at a given pressure the transition temperature, 
 a detailed thermodynamical analysis \footnote{See, for instance, for 
the case of water Chapt. $10$ of Ref\cite{Preparata}} is necessary, 
however one notes that the gaseous state is certainly stable if $F_{i}е$ 
given by Eq.(\ref{eq:3}) 
is such that

\begin{equation}
F_{i} \geq 1 
\label{eq:4} 
\end{equation}

    \item
(v)\ for negligible $F_{i}е$ the equilibrium density is determined by

\begin{equation}
p = p_{_{Pauli}}+p_{c} \ \ ;
\label{eq:5} 
\end{equation}

where p is the external pressure, $p_{_{Pauli}}е$ is the \textit{Pauli 
pressure} given by

\begin{equation}
p_{_{Pauli}}=-\frac{\partial}{\partial V}\left( 2V\int_{0}^{p_{F}е}е\frac{d^{3}p}{(2\pi)^{3}}\frac{p^{2}}{2m} \right)_{N}=
\frac{2}{3}b\rho^{\frac{5}{3}} \ \ \ \ \ \ \ \ \ \  \left(b=\frac{\pi^{3}}{10m}\right)  ;
\label{eq:6} 
\end{equation}

and $p_{c}е$ is the \textit{coherent (negative) pressure}

\begin{equation}
p_{c}=-\frac{\partial}{\partial 
V}(N\Delta)_{N}=-\frac{a}{3}\rho^{\frac{4}{3}} \ \ \ \ \ \ \ \ \  \left( a=2.95e^{2}е\right)е .
\label{eq:7} 
\end{equation}
\end{description}е
With these basic results let us turn to the Sun's parameters predicted 
by the SSM, reported in Table I. Our analysis, admittedly coarse and 
preliminary, will simply consist in computing for each solar radius r 
the gap $\Delta$ and the incoherent fraction $F_{i}е$ in terms of the 
temperature T and the density $\rho$ and check whether the condition 
(\ref{eq:4}) 
is satisfied.

When this happens the gaseous state is stable and the evaluation of 
the SSM reliable. If on the other hand $F_{i}(r)\leq 1$, according to 
our theory (and our rough approximation) the basic assumption of the 
SSM fails, for the plasma finds it energetically favorable to collapse 
to the \textit{Coherent Ground State} (CGS). \par
We emphasize that, though coarse, our analysis is basically correct, 
for the gaseous state comprises the exterior solar shell and the SSM 
solution \textit{conquers the interior} starting from the known exterior 
structure through the basic equilibrium equations, and it will thus 
break down when the equation of state hypothesis breaks down. On the 
right columns of Table I we have reported our predictions for the 
gap $\Delta$ (Eq.( \ref{eq:2})) and the incoherent fraction $F_{i}$ 
(Eq. (\ref{eq:3})) for the pressure and temperature of the SSM for the 
electrons' plasma. \par

Proceeding from the outside in, we see that the SSM is stable until 
$\frac{r}{r_{\odot}} \approx 0.3$, after which value the incoherent 
fraction becomes smaller than $1$, and a condensed coherent phase 
gets formed. Thus we conclude that for $\frac{r}{r_{\odot}}\leq 0.3$ the 
SSM breaks down and a CGS appears. The question now is what are its 
features ? \par
The answer can immediately be found through the pressure equation 
(\ref{eq:5}). Taking the SSM value $p=p(0.3)=9.3*10^{9}$ bar, we find for 
the equilibrium density

\begin{equation}
\rho_{eq} = 1.29\cdot 10^{27}cm^{-3} \ .
\label{eq:8} 
\end{equation}

At this density the gap is

\begin{equation}
\Delta_{eq} = 5.9 \ keV \ ,
\label{eq:9} 
\end{equation}

and $(F_{i})_{eq}$ utterly negligible if we take the temperature T, as 
we must, equal to that predicted by the SSM at $\frac{r}{r_{\odot}}\simeq 
0.3$, i.e.

\begin{equation}
T(0.3) = 0.55 keV = 6.41 \cdot 10^{6} \ K \ .
\label{eq:10} 
\end{equation}

We are now in a position to give a preliminary answer to the question 
of the title of this paper. According to QED at the center of the Sun 
there hides a coherent state of the electrons' plasma whose density 
and temperature are given by (\ref{eq:8}) and (\ref{eq:10}) respectively. 
On the other hand, at the temperature (\ref{eq:10}) the plasma of 
protons and of the other nuclei that are produced through the 
reactions of nuclear fusion are predicted to be in the gaseous 
state\footnote{This can easily be appreciated by noting that their 
incoherent fractions exceed that of the electrons by the large mass 
ratios.}. The size of such \textit{coherent core} can be easily estimated 
by noting that according to the SSM it comprises $0.65$ of the mass of 
the Sun, or a number of nucleons $N_{N}=7.8\cdot 10^{56}$, thus 
$(r_{\odot}=7\cdot 10^{10} \ cm)$

\begin{equation}
\frac{r}{r_{\odot}}=\left[\frac{N_{N}}{\rho}\frac{3}{4\pi}\right] ^{1/3} 
\frac{1}{r_{\odot}}=0.075 \ .
\label{eq:11} 
\end{equation}

One of the most important (preliminary) consequences of these uncanny 
results is upon the solar neutrino problem, that hinges mainly on the 
high energy neutrinos emerging from the $\beta$ - decay of $B^{8}$.
The temperature dependence of their flux $\Phi(B^{8})$ is very strong 
\cite{Bahcall}:

\begin{equation}
\Phi(B^{8})\approx const \cdot T^{18} \ ,
\label{eq:12} 
\end{equation}

thus changing the central temperature from $T(0) = 1.54 \cdot 10^{7} \ K$ 
to the value (\ref{eq:10}) entails a suppression $\left( 
\frac{0.64}{1.54}\right)^{18} = 1.4 \cdot 10^{-7}$, even though some 
enhancement of the constant in (\ref{eq:12}) is to be expected due to 
the increased central density (see Table I and (\ref{eq:8})). It has 
been remarked\footnote{See the book of Ref. \cite{Bahcall}, pag.139 .} that a complete 
suppression of the $B^{8}$-neutrinos would (a) solve the solar 
neutrino problem, and (b) have no consequence on the Sun's internal 
structure: the $\textit{coherent core}$ is thus seen to pass brilliantly 
this test. But how about the numerous hurdles that the SSM overcomes in 
an apparently natural way ? It is too soon to tell, a theoretical 
effort is needed comparable to the one carried out within the SSM, 
that we hope people will feel motivated to reproduce. \par
In conclusion, we have submitted the SSM and its equation of state to 
a critical analysis within the framework of $\textit{QED Coherence}$ 
\cite{Preparata}. We have found that for $\frac{r}{r_{\odot}} \leq 0.3$ 
the SSM is unstable against a phase transition towards the CGS of the 
electrons' plasma. As a result a dense, \textit{cold coherent core} is 
predicted to arise characterized by Eqs (\ref{eq:8})-(\ref{eq:10}), 
whose first recognizable and observable consequence is the suppression 
of $B^{8}$-neutrinos, and the solution of the solar neutrino problem. 
\par
A last intriguing remark. In the formation of the \textit{coherent core} 
a large energy is released:

\begin{equation}
E_{c} = N_{N}\left( 
\Delta - \frac{\pi^{3}}{10m}\cdot \rho^{2/3}\right) = 3.7 \cdot 
10^{48} \ erg 
\label{eq:13} 
\end{equation}

in a short time, the amount that the Sun at its present rate releases 
during $30$ million years! It is a tantalizing thought that this 
monstruous explosion be at the origin of the solar system. \par
I wish to thank E. Del Giudice for interesting conversations.

\newpage

\begin{table}
  \begin{tabular}{|c|c|c|c|c|c|c|c|c|}
	\hline
	$r\backslash r_{\odot}$ & $m\backslash m_{\odot}$ & P (bar) & T (keV) & 
	$\rho (cm^{-3})$ & $\Delta (keV)$ & $F_{i}$ & $-p_{c} (bar)$ & $p_{_{Pauli}} 
	(bar)$  \\
	\hline
	\ 0.00\  & \ 0\ & \ $2.4\cdot 10^{11}$\  & \ 1.327\  & \ $9.18 \cdot 
	10^{25}$\  & \ 2.66\  & 
	\ 0.35\  & \ $1.1 \cdot 10^{11}$\ & \ $4.9\cdot 10^{10}$\  \\
	\hline
	\ 0.06\  & \ 0.10\  & $1.9 \cdot 10^{11}$ & \ 1.246\  & \ $6.64\cdot 
	10^{25}$\  & 
	\ 2.39\  & \ 0.44\ & $7.7 \cdot 10^{10}$\ & \ $2.55\cdot 10^{10}$  \\
	\hline
	\ 0.12\  & \ 0.12\  & \ $1.1\cdot 10^{11}$\  & \ 1.034\  & \ $4.34 
	\cdot 10^{25}$\  & \ 2.07\  & \ 0.45\  & \ $4.4\cdot 10^{10}$\  & \ $1.4\cdot 10^{10}$\   \\
	\hline
	\ 0.16\  & \ 0.23\  & \ $7.0\cdot 10^{10}$\  & \ 0.902\  & \ $3.03 
	\cdot 10^{25}$\  & \ 1.84\  & \ 0.48\  & \ $2.69\cdot 10^{10}$\  & \ 
	$7.7\cdot 10^{9}$\   \\
	\hline
	\ 0.20\  & \ 0.34\  & \ $4.4\cdot 10^{10}$\ & \ 0.798\  & \ $2.12 
	\cdot 10^{25}$\  & \ 1.63\  & \ 0.54\  & \ $1.68\cdot 10^{10}$\  & \ 
	$4.3\cdot 10^{9}$\   \\
	\hline
	\ 0.26\  & \ 0.53\  & \ $1.8\cdot 10^{10}$\ & \ 0.641\  & \ $1.07 
	\cdot 10^{25}$\  & \ 1.30\  & \ 0.69\  & \ $6.75\cdot 10^{9}$\  & \ 
	$1.4\cdot 10^{9}$\   \\
	\hline
	\ 0.31\  & \ 0.65\  & \ $9.3\cdot 10^{9}$\ & \ 0.552\  & \ $6.36 
	\cdot 10^{24}$\  & \ 1.09\  & \ 0.89\  & \ $3.37\cdot 10^{9}$\  & \ 
	$5.7\cdot 10^{8}$\   \\
	\hline
	\ 0.43\  & \ 0.83\  & \ $1.8\cdot 10^{9}$\ & \ 0.398\  & \ $1.72 
	\cdot 10^{24}$\  & \ 0.71\  & \ 1.85\  &  & е  \\
	\hline
	\ 0.49\  & \ 0.89\  & \ $7.7\cdot 10^{8}$\ & \ 0.306\  & \ $8.58 
	\cdot 10^{23}$\  & \ 0.56\  & \ 2.84\  & е & е  \\
	\hline
	\ 0.58\  & \ 0.94\  & \ $2.3\cdot 10^{8}$\ & \ 0.265\  & \ $3.21 
	\cdot 10^{23}$\  & \ 0.40\  & \ 4.97\  & е & е  \\
	\hline
	\ 0.68\  & \ 0.97\  & \ $7.1\cdot 10^{7}$\ & \ 0.205\  & \ $1.30 
	\cdot 10^{23}$\  & \ 0.30\  & \ 7.36\  & е & е  \\
	\hline
	\ 0.83\  & \ 0.99\  & \ $8.2\cdot 10^{6}$\ & \ 0.094\  & \ $3.35 
	\cdot 10^{22}$\  & \ 0.19\  & \ 4.76\  & е & е  \\
	\hline
	\ 0.91\  & \ 0.999\  & \ $1.2\cdot 10^{6}$\ & \ 0.059\  & \ $1.04 
	\cdot 10^{22}$\  & \ 0.13\  & \ 5.44\  & е & е  \\
	\hline
	\ 1.00\  & \ 1.00\  & \ 0.12\ & \ $4.9\cdot 10^{-4}$\  & \ $1.81 
	\cdot 10^{17}$\  & \ $2.73 \cdot 10^{-3}$\  & \ 2.55\  & е & е  \\
	\hline
  \end{tabular} 
  \vskip 3 cm
\caption{The SSM predictions for mass, pressure, temperature and 
density at different depths. The last four columns report the 
predictions of \textit{QED coherence} for the gap, the incoherent 
fraction, the coherent (negative) pressure and the Pauli pressure of 
the plasma of the electrons.}
\end{table}

\end{document}